\documentclass[12pt]{article}
\usepackage[utf8x]{inputenc}
\usepackage{amssymb}
\usepackage{amsmath}
\usepackage{epsfig}
\usepackage{graphicx}
\usepackage{caption}
\usepackage{subcaption}
\usepackage{color}
\usepackage{ulem}

\begin{document}

\begin{titlepage}
\vskip1cm   
\begin{flushright}
\end{flushright}
\vskip1.25cm
\centerline{\bf \large Multi-Scale Distributed Representation for  Deep Learning 
} 
\centerline{\bf \large and its Application to b-Jet Tagging} 
        
\vskip1cm \centerline{ 
Jason Lee,   Inkyu Park,  Sangnam Park$^*$}
\vspace{1cm}
\centerline{ Department of Physics,
University of Seoul, Seoul 02504 \rm KOREA}
 \vspace{0.1cm}
 \centerline{
\tt{(jason.lee@uos.ac.kr,icpark@uos.ac.kr,u98parksn@gmail.com)}} 
 
  \vspace{1cm}
{ 
Recently machine learning algorithms  based on 
deep layered artificial neural networks (DNNs)  have been applied to a wide variety of high energy physics problems such as jet 
tagging or event classification. 
We explore a simple but effective preprocessing step which transforms each real-valued observational quantity or input feature into a binary number with a fixed number of digits. Each binary digit 
represents the quantity or magnitude in different scales. We have shown that this approach  improves the performance of DNNs significantly for some specific tasks without any further complication in feature engineering. We apply this multi-scale distributed binary representation to deep learning 
on b-jet tagging using daughter particles' momenta and vertex information.
}

\vspace{1cm}
\noindent
PACS numers: 13.90.+i, 13.87.-a \\
Keywords: Machine Learning, Jet Tagging 
\end{titlepage}

\section{Introduction}
In the most recent machine learning, 
deep-layered artificial neural networks (DNNs) have been successfully applied to wide range of physics problems from phase transitions in statistical mechanics \cite{Evert:2017nature,Juan:2017nature} to quark/gluon jet discrimination in high energy physics \cite{Patrick:2017jhep}. However, since there are no comprehensive rules or principles by which we can select a particular architecture or the input features of the network model for the corresponding tasks, the so called hyper-parameters determining the architecture of the networks and the overall learning processes are examined by pain-stacking trials and errors for a given set of input features.
The resulting parameters for the networks are believed to be optimal, but they are too specific to be applied for other similar tasks. So for these similar tasks, trials and errors often need to be repeated.

\begin{figure}
        \centering
        \begin{subfigure}[b]{0.4\textwidth}
                \centering
                \includegraphics[width=\textwidth]{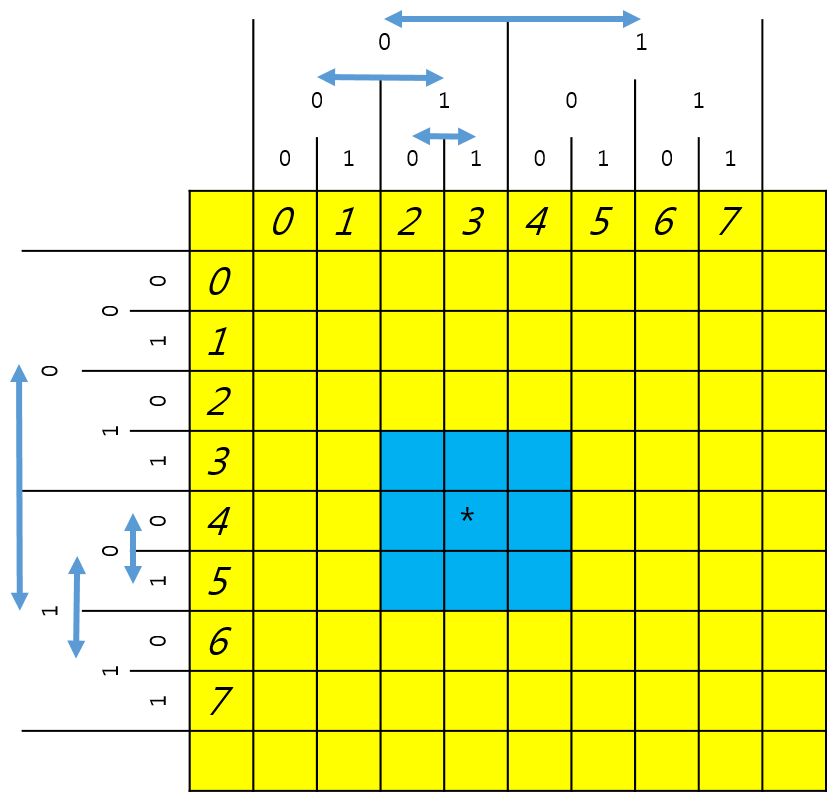}
                \caption{}
        \end{subfigure}%
        ~
        \begin{subfigure}[b]{0.4\textwidth}
                \centering
                \includegraphics[width=\textwidth]{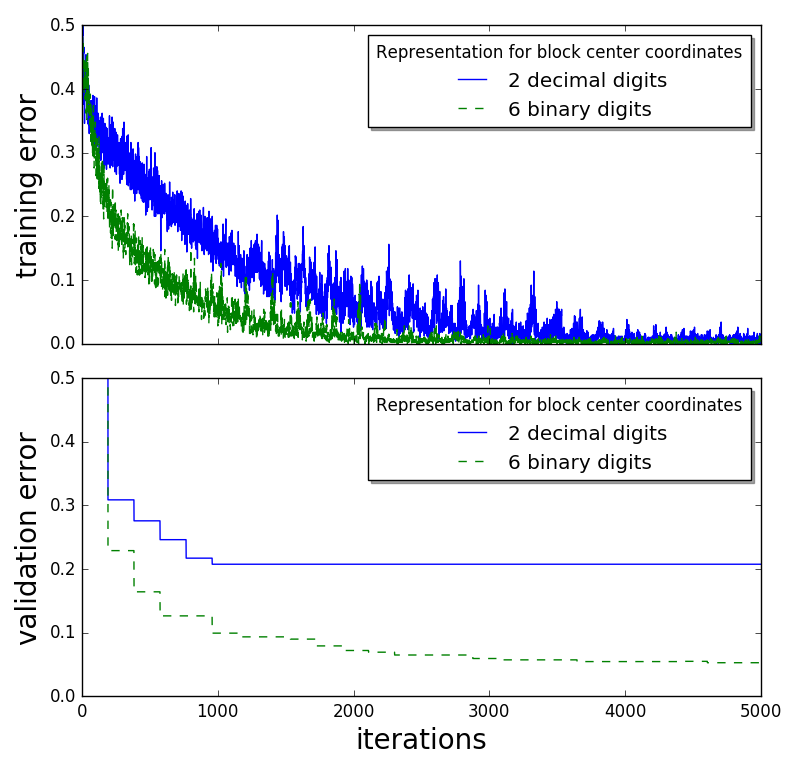}
                \caption{}
        \end{subfigure}%

        
\caption{\label{SpinBlock} \footnotesize
Multi-scale representation of site position distributed in 6 binary digits: the left most digits represent the region in largest scale.
}
\end{figure}

However, if we use good input features or representations for DNNs, we can reduce the repetitions, since the good input features help DNNs to learn good internal representations \cite{Yann:2015nature,Yoshua:2013ieee} with less dependence on its detailed architectures. 
Our investigation stems from a following simple question: what happens if one uses 
multiple variables with smaller dynamic range instead of one variable with large dynamic range as an input feature? 
We have tested this 
with a simple model. We designed a network to predict the sign of the sum of the spin states on a $3\times3$ block within a $10\times10$ lattice of random spins provided with the coordinates of center of the block limited within inner $8\times8$ lattice as shown in Fig.~\ref{SpinBlock}. The network had to find out what the uniformly distributed input variables for coordinates of block center site meant and how to process them to perform the task, as this was not coded in explicitly. In this setting, when we provided the network with a binary  representation of the site coordinate, for example ($011$,$\,100$) instead of ($3$,$\,4$), the network learned more quickly and its training and test error were less as shown on the Fig.~\ref{SpinBlock}b. 
We can consider that this phenomenon is caused by the increased sparsity resulting from transforming the input values from a large base range to a smaller base (see Fig.~\ref{sparsityPlot} for the related discussion), having a distributed representation \cite{Yann:2015nature,Yoshua:2013ieee} simply implemented with a number of binary digits for each input value in this case. 

 Inspired by these findings, we transformed each real-valued feature of jet constituents to a $k$-number of binary digits as the input features of deep neural networks for b-jet tagging \cite{bjet} to test whether this simple preprocessing can improve the performance of the networks on a harder problem. We could implement the multi-scale distributed (MSD) representation in various ways, but the representation with $k$-number of binary digits (MSD$_2$ $k$-digits) is the simplest and convenient one. Thus we studied only this binary implementation,
 which shall be denoted as 
 MSD$_2$ $k$-digits or simply MSD $k$-digits from now on.   

In section 2, we will describe briefly data set used in this work, before exploring the detailed preprocessing of jet data and MSD digit representation. Section 3 covers the network architectures and its learning processes, and section 4 presents the results followed by our concluding remarks.


\section{Jet Data and MSD $k$-digit Representation}\label{Sec2}
To test the improvement of the MSD representational algorithm against the typically used real-valued ones on b-jet tagging, we generated $t\bar{t}$ events in $pp$ collisions at $\sqrt{s}=13$ TeV using the next leading order POWHEG \cite{Oleari:2010nx} event generator matched to PYTHIA8 \cite{Pythia} for hadronization. Jets with transverse momentum $p_T > 30$ GeV and pseudo rapidity $|\eta|<2.4$ with at least two daughter constituents are selected. We used FastJet \cite{Fastjet} with the anti-$k_T$ algorithm \cite{anti-kt} with a jet radius $R=0.4$ for jet finding and clustering. We used Delphes for fast detector simulation \cite{Delphes}. We did not separated samples by $p_T$ range of jet for quick and simple test of MSD representation.

We used only jets initiated from light partons such as up, down, strange quarks and gluon (light-jet) for background. 
The number of b-jets and light-jets used in this study are the same, each with $70k$ samples. The total $140k$ jet samples are divided into a training set ($\mathcal{D}^{train}$) of $100k$ samples, with a early validation set ($\mathcal{D}^{validation}$) of $10k$ samples for early-stop to be collectively applied to all member networks 
and test set ($\mathcal{D}^{test}$) of $30k$ samples for the final performance evaluation.

\begin{figure}[t!]
\begin{center}
\includegraphics[width=16cm,clip]{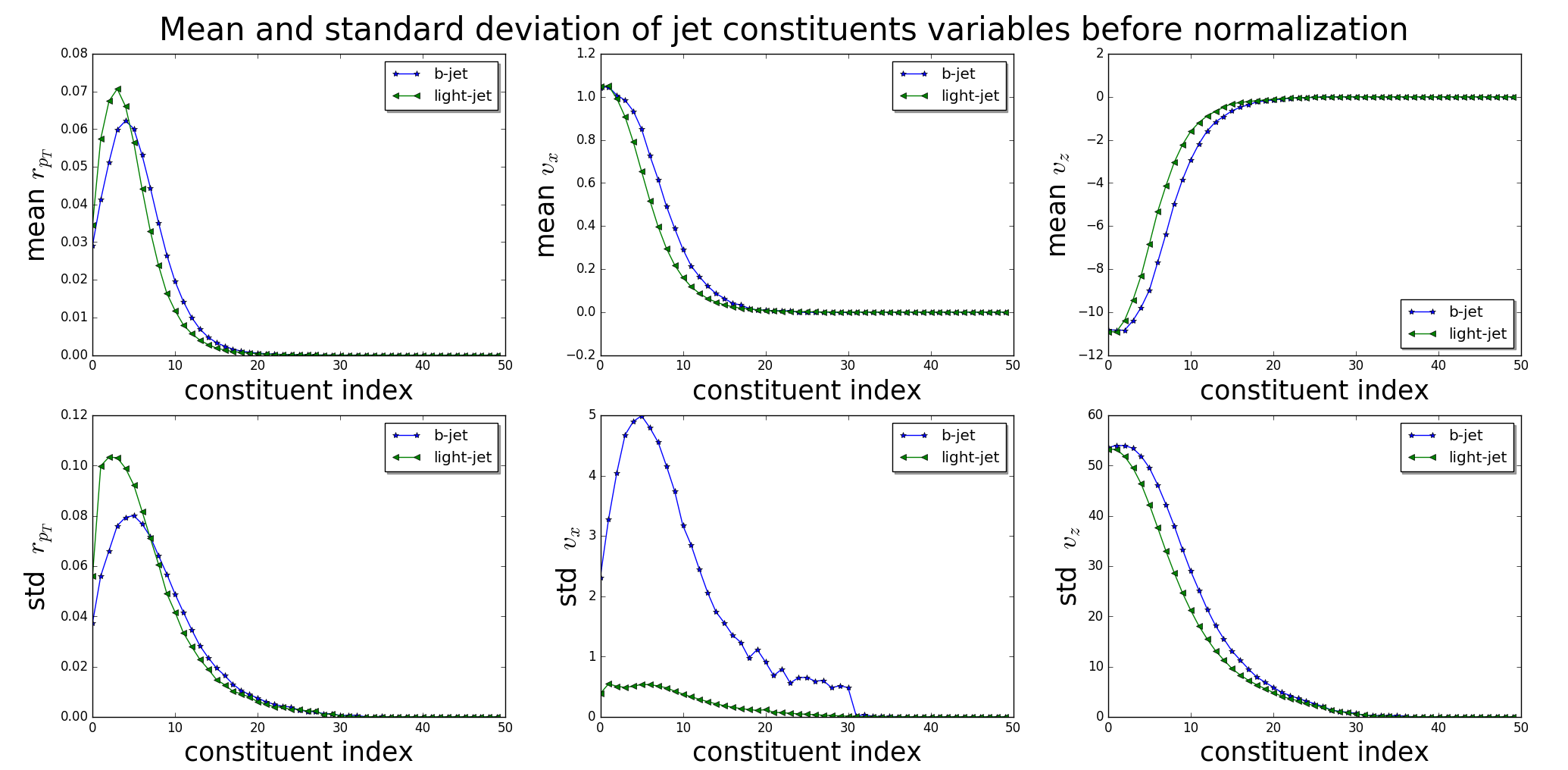}
\includegraphics[width=16cm,clip]{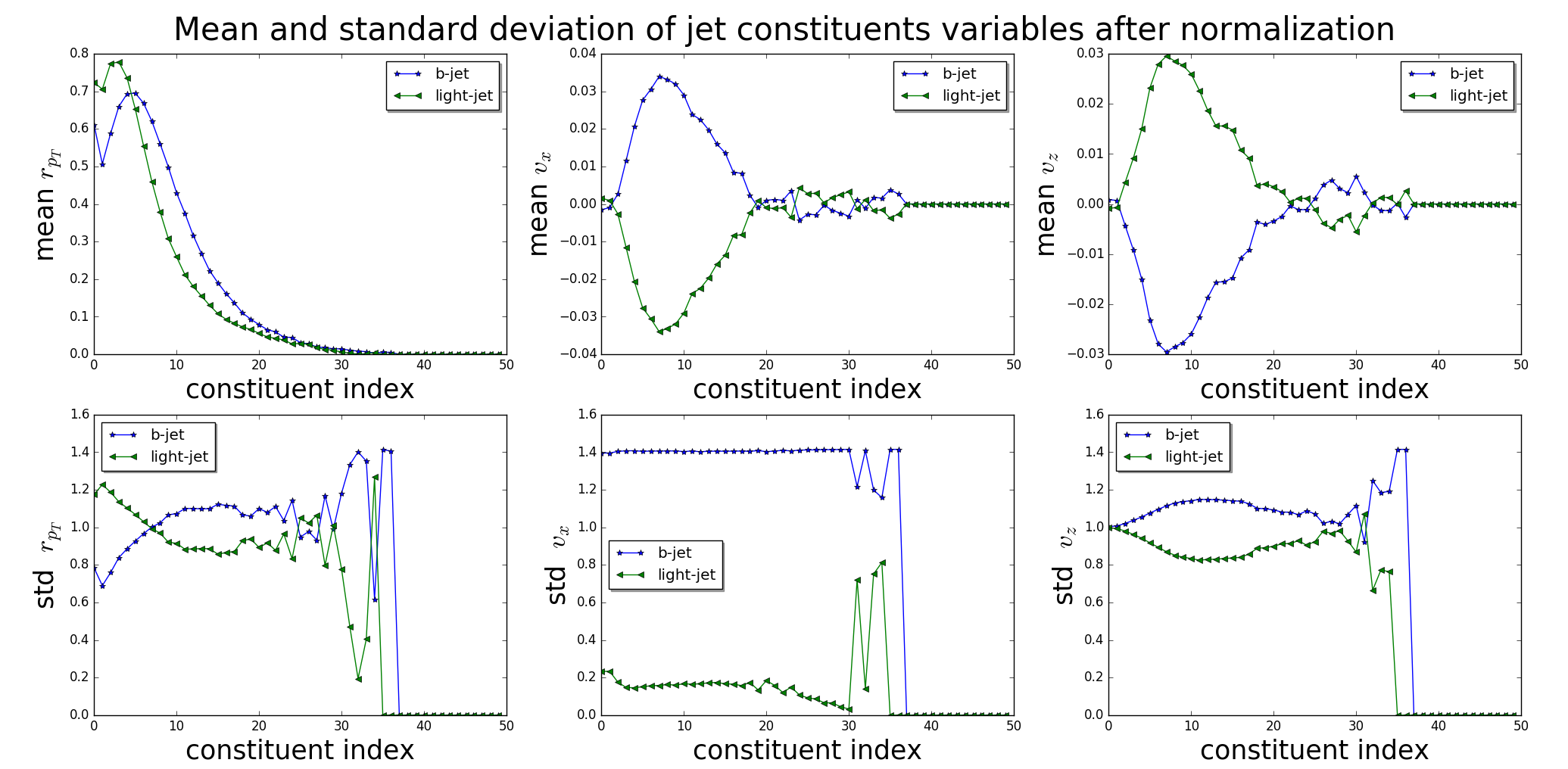}
\end{center}
\caption{
\label{trackPlots} \footnotesize
Mean and standard deviation (std) plots for jet constituents $r_{P_T},v_x$,$v_z$  along constituent index, before (upper two rows) and after (lower two rows) normalization.
}
\end{figure}

\begin{figure}[htb!]
\begin{center}
\includegraphics[width=15cm,clip]{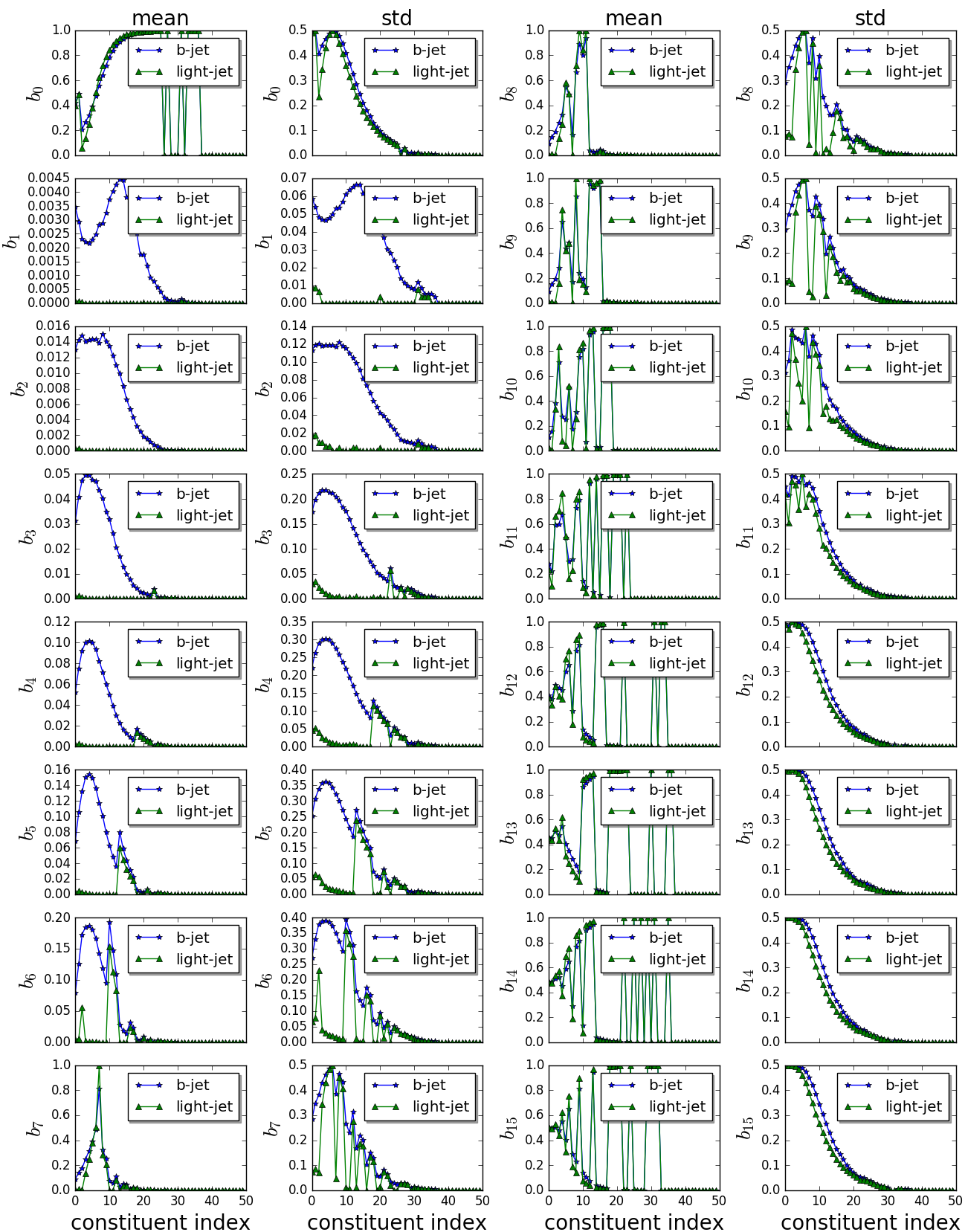}
\end{center}
\caption{
\label{BinaryTrackPlots} \footnotesize
Mean and standard deviation plots for MSD$_2$ $16$-digits on $v_x$ of jet constituent  in simulated samples: each horizontal axis represents constituent particle whereas each vertical axis is for mean or standard deviation (std).  
}
\end{figure}

We used the jet constituent's relative transverse momentum ratio and it's vertex positions as input features ($r_{p_T}$,$v_x$,$v_y$,$v_z$). Jets have different number of its constituents or particles. We used a fully connected feed forward networks which take a fixed number of variables for its input, so we truncated the sequence of jet constituents variables up to the $50$th constituent.
And if the number of jet constituents $n$ is less than $50$, we set rest values to zeros. So, $\mathbf{x} = (x^1,x^2,x^3,x^4,\ldots,x^i,\ldots,x^{199},x^{200})$ represents the following data structure,
\begin{displaymath}\label{pre-MSD}
\mathbf{x} = \left\{ \begin{array}{ll}         (r_{p_T}^1,v_x^1,v_y^1,v_z^1,\ldots,r_{p_T}^n,v_x^n,v_y^n,v_z^n,\ldots,r_{p_T}^{50},v_x^{50},v_y^{50},v_z^{50})
& \textrm{if $n \ge 50$} \\
(r_{p_T}^1,v_x^1,v_y^1,v_z^1,\ldots,r_{p_T}^n,v_x^n,v_y^n,v_z^n,0,0,0,0,\ldots,0,0,0,0)                    
& \textrm{if $n < 50$}
\end{array} \right.
\end{displaymath} 
where upper index represents the order of jet constituent in some rule, e.g. ordering by $p_T$.

Before transforming the sequence of jet constituents variables $\mathbf{x}$ to MSD $k$-digit representation, zero-centering and normalization processes are performed to adjust the dynamic ranges of each component. This data-oriented processed representation will be called real-valued representation and it is defined as follows:

\begin{align}\label{normalization}
 u^i = \left\{\begin{array}{ll}
             \frac{x^i}{\sigma^i}  & \textrm{if $x^i$ is momentum weight,}\\
             \frac{x^i-\langle x^i\rangle}{\sigma^i} & \textrm{if $x^i$ is vertex position.}
             \end{array}\right.
\end{align}
Fig.~\ref{trackPlots} shows distributions of ratio $r_{p_T}$ of constituent's $p_T^{const.}$ to jet  $p_T^{jet}$, constituent $v_x$, and $v_z$ before and after the normalization 
process~(\ref{normalization}).

\begin{figure}[ht!]
\begin{center}
\includegraphics[width=14cm,clip]{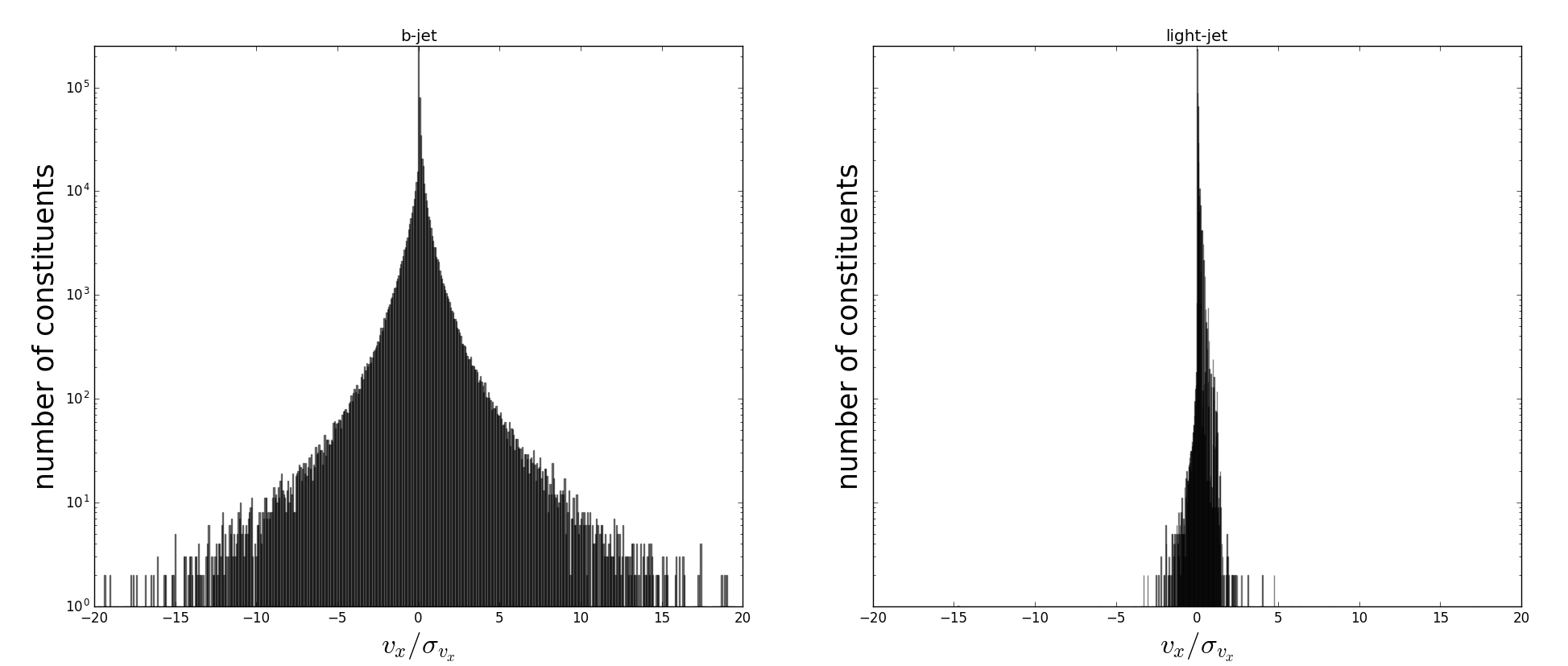}
\includegraphics[width=14cm,clip]{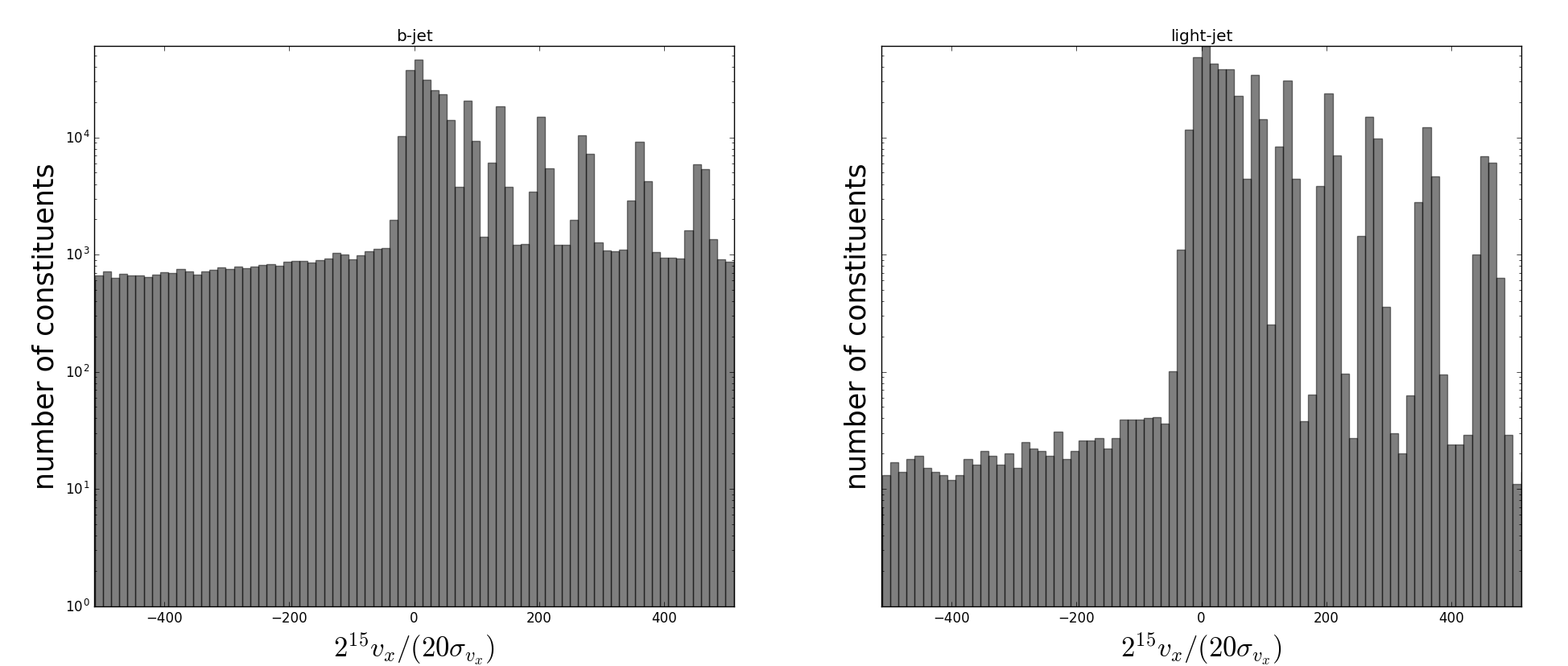}
\end{center}
\caption{
\label{vx_distribution} \footnotesize
The distribution of jet constituents' $v_x$  in simulated samples: (top) log-scale vertex position significance $v_x/\sigma_{v_x}$ and (bottom) digitized vertex position with interval $(20\sigma_{v_x})/2^{15}$ near zero. The plots on the left are for the b-jet whereas the right ones  for the light-jet. 
}
\end{figure}

Transforming the real-valued representation, $\mathbf{u}$, to MSD$_2$ $k$-digit representation can be simply performed by converting decimal numbers to binary digits with signed magnitude representation after clipping and rounding process as follows:

\begin{align}\label{MSDprocess1}
z^i = \left\{ \begin{array}{ll}
               2^{k-1}-1                   & \textrm{if $\frac{u^i}{\epsilon} \ge 2^{k-1}-1$}\\
               \lfloor \frac{u^i}{\epsilon} \rceil  & \textrm{if $-2^{k-1}+1 < \frac{u^i}{\epsilon} < 2^{k-1}-1$}\\          
              -2^{k-1}+1                   & \textrm{if $\frac{u^i}{\epsilon} \le -2^{k-1}+1$,}
             \end{array} \right.
\end{align}
and then the resulting decimal number $z^i$ is represented as a signed binary $B^i$
\begin{align} \label{MSDprocess2}
z^i \rightarrow B^i = (b_0^i,b_1^i,b_2^i,\ldots,b_{k-1}^i) 
\end{align} 
where $i=1,2,\ldots,200$ and $b\in\{0,1\}$. The number of features are increased to $200 \times k$ by the transformation. While the optimal interval or resolution $\epsilon$ may need systematic analysis, we just set it by $20\sigma / (2^{k-1}-1)$, with $\sigma =1$ for the real-valued representation. Fig.~\ref{BinaryTrackPlots} shows the mean and standard deviations of the vertex position, $v_x$, for each of the digits in the MSD$_2$ $16$-digit representation. The first digit, $b_0$, showing the sign bit, followed by $b_1$ for the first significant figure, to the last digit $b_{15}$, are the so called {\sl signed} MSD$_2$ $16$-digit representation\footnote{Let $B$ a $k-1$ digit binary number. Its signed $k$-digit representation is defined as follows: 
For 
negative $B$, 
one has the binary representation of $2^{k-1}+|B|$ while, for non negative $B$,  one simply adds an extra $0$ in front of $B$ to get $k$-digit
representation. On the other hand, for its 
two's complement representation below, one has $k$-digit binary representation of $2^k-|B|$ for negative $B$ whereas the definition of  $k$-digit representation remains the same for non negative $B$. 
}.
The horizontal axis represents the jet constituent index. The plots of the mean values of the most significant binary digits are smooth and it starts to be more discontinuous and the digit significance decreases.
The reason for such discontinuity is due to the presence of sharp peaks near zero in the distribution of simulated $v_x$, as seen in Fig.~\ref{vx_distribution}.

\section{Networks Architectures and Learning processes}\label{Sec3}

\begin{figure}[ht!]
\begin{center}
\includegraphics[width=12cm,clip]{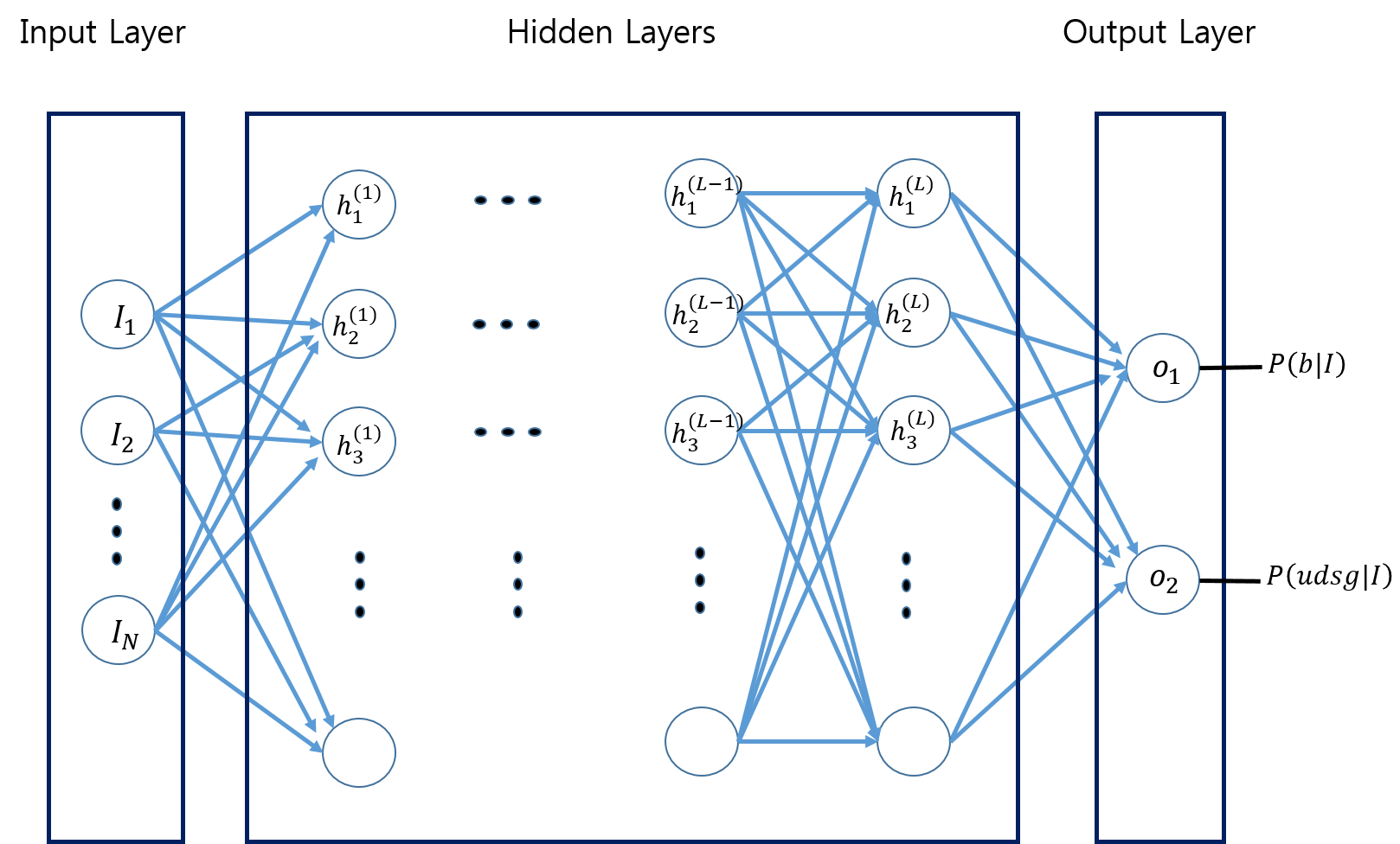}
\end{center}
\caption{\label{architecture} \footnotesize Schematic architecture of fully connected neural network (FCN) used for b-jet tagging with jet constituents variable in real-valued or MSD $k$-digit representation. Blue lines with arrow connecting units of lower and higher layers represent the direction of information flow in which a signal from one unit in a  layer enters into all units in its next layer. 
}
\end{figure}

In this study, we used a fully connected neural network (FCN) for b-jet tagging. The schematic architecture of this network is shown in Fig.~\ref{architecture}. 
We used the rectified linear unit (ReLU) activation function \cite{relu} for all units $h^{(l)}_i$'s in each hidden layer indexed by the superscript $l$ and the $softmax$ function for two output units $o_1$ and $o_2$.
The categorical cross-entropy is used as the cost function for the training.

Below are input units for each of the real-valued jet constituent features and its corresponding MSD $16$-digit representation.
\begin{eqnarray}
\label{inputUnit}
  \phantom{aaaaaaaaa} I^{real}_i &=& u^i \\
     I^{MSD}_{j} &=& b^{i}_{m},    \quad\quad\quad\quad j = 16\times(i-1) + m + 1
\end{eqnarray}
where binary position index $m = 0,\ldots,15$  and $i = 1,\ldots,4\times50$ for the real-valued input feature. 
We prepared both ``Deep" and ``Not so deep" versions for each input representation to check the stability of our result.
where binary position index $m = 0,\ldots,15$  and $i = 1,\ldots,4\times50$ for the real-valued input feature. 
We prepared both ``Deep" and ``Not so deep" versions for each input representation to check the stability of our result.

Neural networks were trained using the Theano Python library \cite{theano} on GPUs using the NVIDIA CUDA platform. 
The connection weights of the networks were initialized with the Xavier Initialization scheme \cite{xavier_initialization}. The Adam \cite{adam} algorithm was used to update the weights up to each early-stop epoch specified in Table.~\ref{network_summary}, using a batch size of 1000 which corresponds to 100 iterations per epoch in the training stage.

\begin{figure}[ht!]
\begin{center}
\includegraphics[width=15cm,clip]{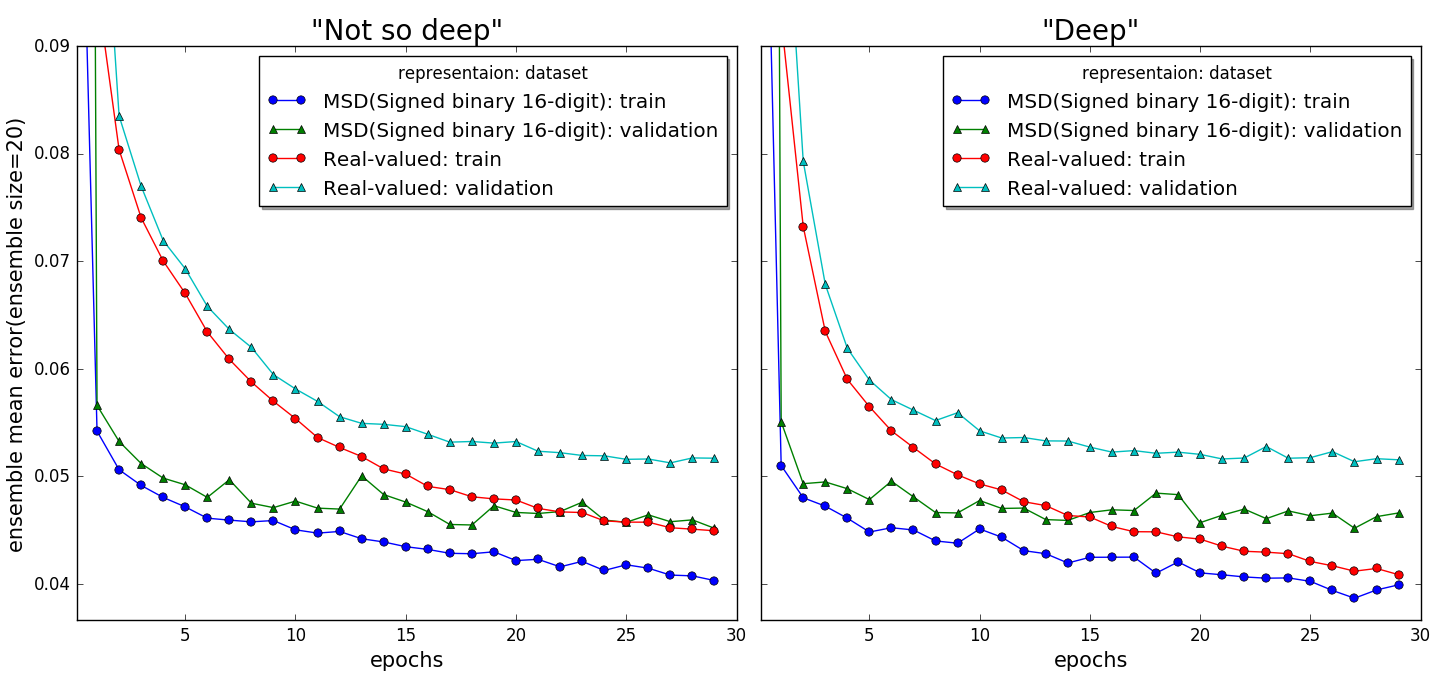}
\end{center}
\caption{
\label{learning_progress} \footnotesize
Learning progress (mean of training and validation errors by member networks of ensemble in try-out phase with ensemble size $=20$ along iterations.
Two curves showing higher error rates at early epochs are for the real-valued representation on each ``Not so deep" and ``Deep" plots, while the lower two curves for the MSD representation. Curves on training error are lower than those of validation for each representation. 
}
\end{figure}

\begin{table}[htb]
\begin{center}
\begin{tabular}{c|cc}
Hyper-Parameters                      & ``Not so deep"            & \quad ``Deep"\\
\hline
$N^{hidden}_{layer}\times N^{hidden}_{unit}$,~activation & 3$\times 1000$,~ ReLU    & 6$\times 500$,~ ReLU\\
Early stop epoch for MSD-16        & 30                       &  25 \\
Early stop epoch for Real-valued   & 25                       &  25 \\
Ensemble Size                      & 100                      &  100 \\
$L_2$ penalty parameter for MSD-16                   & 0.001                    & 0.001 \\
$L_2$ penalty parameter for Real-valued              & 0.0001                   & 0.0001 \\
\hline
\end{tabular}
\end{center}\caption{\label{network_summary} \footnotesize
Hyper-parameters on regularization and architecture are presented in this table. While the number of input units of network for real-valued representation is $200$, the number of input units of network for MSD-16 is $200\times16 = 3200$. 
}
\end{table}

For regularization, we used the simple ensemble voting method \cite{ensemble_voting} in which the network ensemble was composed of networks trained with just different random seeds for their initial connection weights and a random sequence of mini-batches up to epoch $t$ like this, 
\begin{align} \label{network_ensemble}
\mathcal{N}(t) =\{ f(\mathbf{I};\mathbf{\theta}_1(t)),\ldots,f(\mathbf{I};\mathbf{\theta}_q(t)),\ldots,f(\mathbf{I};\mathbf{\theta}_S(t))\},
\end{align} 
and the ensemble vote was defined as
\begin{align} \label{network_ensemble_voting}
\mathcal{F}(\mathbf{I}) \equiv \frac{1}{S}\sum_{q=1}^{S}f(\mathbf{I};\mathbf{\theta}_q),
\end{align} 
where $f$ is the network model, $\mathbf{\theta}_q(t)$'s represent the connection weights and biases updated up to epoch $t$, with different initialization and mini-batch sequence $\mathcal{D}_q^{train}(t) = \{d_q(0),d_q(1),\ldots,d_q(t)\}$ by random seed with index $q$, and $S$ is size of the ensemble of networks.
We also used $L_2$ penalty \cite{regularization} and early-stop method together with previously mentioned ensemble method. 
The early stop was determined by inspecting networks learning progress plot obtained by try-out learning up to $30$ epochs (Fig.~\ref{learning_progress}), to find the epoch at which the mean of validation errors by $20$ member networks (try-out ensemble size $S_{try-out} = 20$) become stagnated, with the ensemble mean error as below at epoch $t$,
\begin{eqnarray}
\mathcal{E}^{train}(t) &\equiv& \frac{1}{S}\sum_{q=1}^{S}E_{f_q(t)}[d_q(t)], \\
\mathcal{E}^{validation}(t) &\equiv& \frac{1}{S}\sum_{q=1}^{S}E_{f_q(t)}[\mathcal{D}^{validation}],
\end{eqnarray}
where $E_{f_q}[\mathcal{D}]$ represent prediction error rate by member network $f(\mathbf{I};\mathbf{\theta}_q(t))$ of ensemble  $\mathcal{N}$  for data set $\mathcal{D}$ at epoch $t$.
Table.~\ref{network_summary} shows the summary of the architectures and parameters on regularization.

\begin{figure}[ht!]
        \centering
        \begin{subfigure}[b]{0.55\textwidth}
                \centering
                \includegraphics[width=\textwidth]{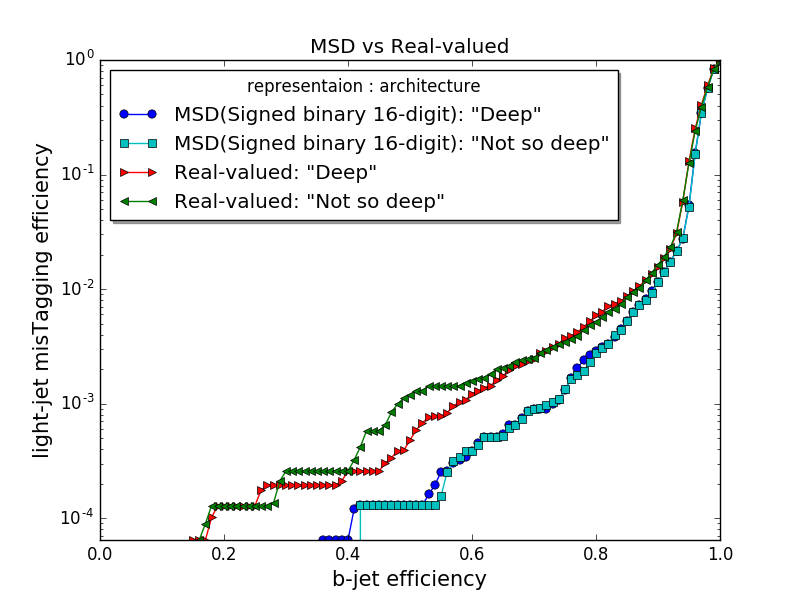}
                \caption{}
        \end{subfigure}%
        
        \begin{subfigure}[b]{0.55\textwidth}
                \centering
                \includegraphics[width=\textwidth]{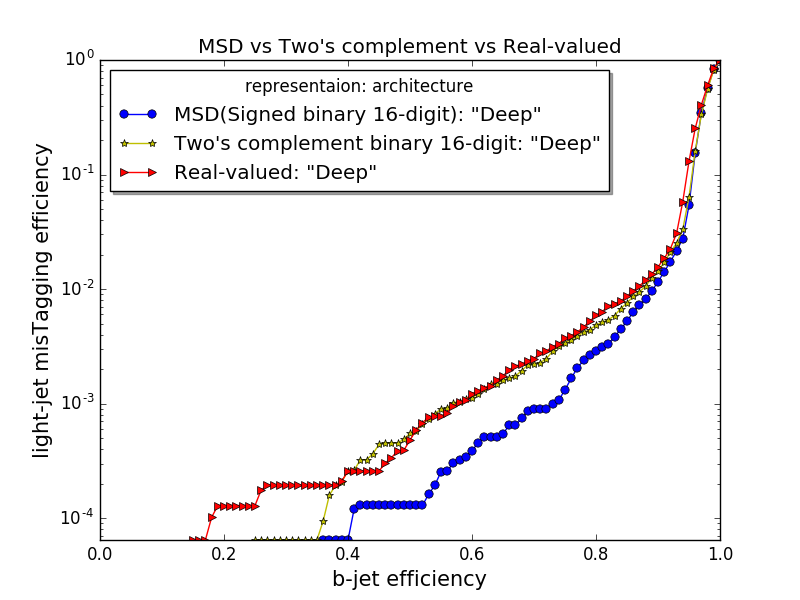}
                \caption{}
        \end{subfigure}%

        
\caption{
\label{Performance} \footnotesize
Performance curves obtained from the output of ensemble vote: the horizontal axis represents b-jet efficiency and the vertical axis represents probability with which the network ensemble identify incorrectly light-jet as b-jet. (a) Plot for comparison of MSD $16$-digit and the real-valued representation. The curves split into two groups, upper two lines for the real-valued and lower two lines for the MSD $16$-digit.   (b) Plot for comparison of MSD $16$-digit, two's complement binary $16$-digit, and the real-valued representations. 
The performance of the  two's complement representation is rather similar to that of the real-valued representation. 
}
\end{figure}

\section{Results}

We obtained performance curves by extracting outputs of the networks, shown in Fig.~\ref{Performance}.
The MSD method shows significant improvement on both ``Deep" and ``Not so deep" networks compared to the real-valued method. 
On the other hand, only a marginal improvement was observed when one uses two's complement binary representation over the real-valued method.
Two's complement representation showed lower performance than the signed MSD representation  as demonstrated in Fig.~\ref{Performance}b.

\begin{figure}[ht!]
\centering
        \begin{subfigure}[b]{0.85\textwidth}
                \centering
                \includegraphics[width=\textwidth]{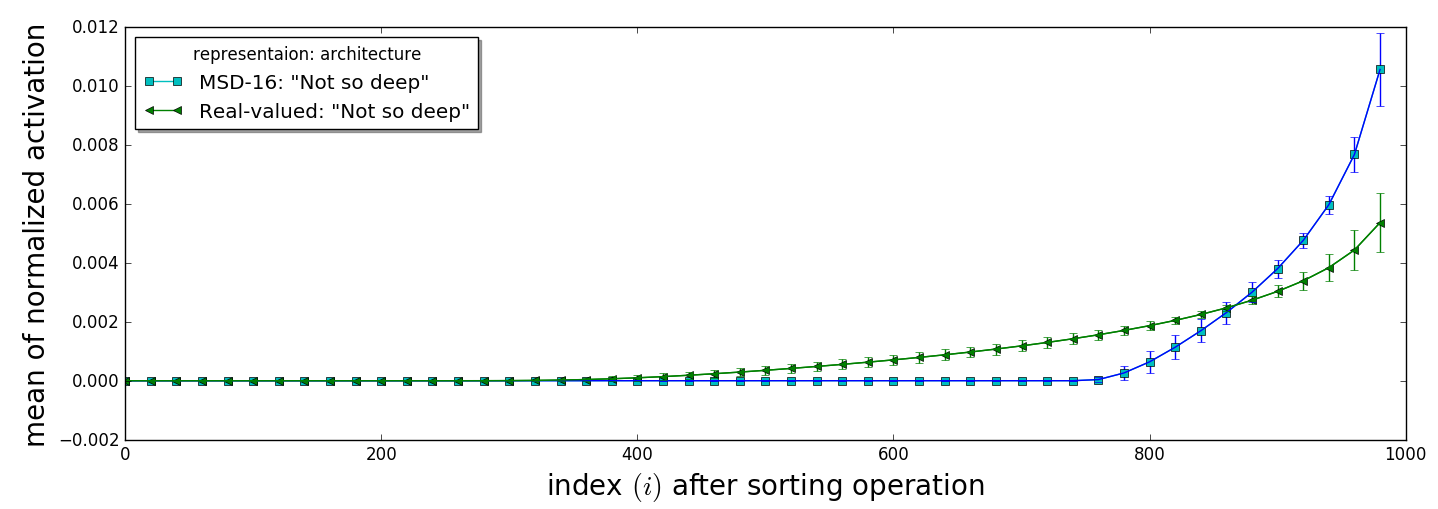}
                \caption{}
        \end{subfigure}%
        
        \begin{subfigure}[b]{0.5\textwidth}
                \centering
                \includegraphics[width=\textwidth]{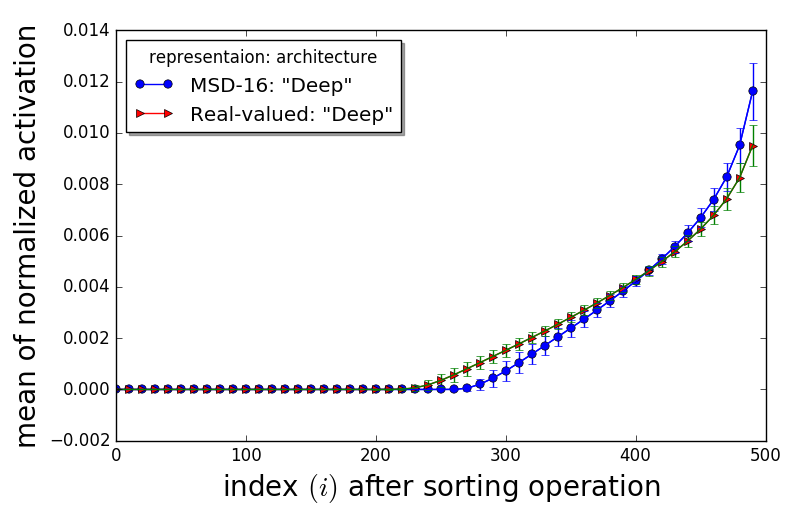}
                \caption{}
        \end{subfigure}%
\caption{
\label{sparsityPlot} \footnotesize
Effect on sparsity: the averages of normalized activation $\langle h^{(1)}_{(i)}/\sum_{i=1}^{N_{h^{(1)}}}h^{(1)}_{(i)}\rangle_{\mathcal{D}^{test}}$ of hidden units in the first layer of  ``Not so deep"  (a)  and  ``Deep" (b) networks for each real-valued  and MSD 16-digit representation to the problem of b-jet tagging, where $h^{(1)}_{(i)}$  is the sorted activation from the smallest to the largest as 
$h^{(1)}_{(1)} \le h^{(1)}_{(2)} \le \ldots \le h^{(1)}_{(N_{h^{(1)}})}$.
}
\end{figure}

There are a few comments on the sparsity of our representation. We found that our MSD representation is more sparse 
than the real-valued representation. This was demonstrated in Fig.~\ref{sparsityPlot} where the averages of normalized activation $\langle h^{(1)}_{(i)}/\sum_{i=1}^{N_{h^{(1)}}}h^{(1)}_{(i)}\rangle_{\mathcal{D}^{test}}$ are plotted. 
In Fig.~\ref{sparsityPlot}a, the blue line is for the MSD representation in the ``Not so deep" network. It clearly shows
that 
the number of deactivated units for this MSD representation is much larger than that for 
the real-valued representation (See the green line). 
Hence we conclude that our MSD representation is more sparse, which partially explains why it shows 
a better performance in this machine learning. Fig.~\ref{sparsityPlot}b is for the ``Deep" networks and basically 
shows the same trends. 

\section{Conclusions}

We demonstrated that a simple transformation from each real-valued input feature to MSD digit representation can lead to a large improvement in the fully connected networks for b-jet tagging without any additional domain specific feature engineering. 


In typical network optimization, one has to examine broad range of hyper-parameters, such as network depth, number of units per layer and regularization parameters.
We have shown that our results from two groups of networks, ``Deep" and ``Not so deep",  are not sensitive to such parameters.

Compared with a typical binary transformation, our MSD conversion described in Eqs. (\ref{MSDprocess1}) and (\ref{MSDprocess2}) from the real-valued feature to signed binary digits reflects the multi-scale property of the original data. For example, typical two's complement representation converts decimal digits $(-3,\,-2,\,-1,\,0,\,1,\,2,\,3)$ to $(101,\,110,\,111,\,000,\,001,\,010,\,011)$, while our signed binary representation converts  to $(111,\,110,\,101,\,000,\,001,\,010,\,011)$. Our choice not only strengthens the multi-scale property but also makes the result information-theoretically effective. For instance, this is because $(111)$, two's complement of $-1$, activates excessive input units compared with the signed representation of $(101)$. 

In this note, we limit our study of the MSD digit representation to the problem of b-jet tagging. However our method can be straightforwardly applied to
many other areas of deep learning problems. Further investigation is required in this direction.

\section*{Acknowledgement}
This work was supported in part by NRF Grant 2017R1A2B4003095.



\begin{thebibliography}{99}

\bibitem{Evert:2017nature} 
  Evert~P.~L.~van~Nieuwenburg, Ye-Hua~Liu \& Sebastian~D.~Huber,
  ``Learning phase transitions by confusion,''
  Nature Physics volume {\bf 13}, 435–439 (2017)

\bibitem{Juan:2017nature}
  Juan~Carrasquilla \& Roger~G.~Melko,
  ``Machine learning phases of matter,''
  Nature Physics volume {\bf 13}, 431–434 (2017)
 
\bibitem{Patrick:2017jhep}
  P.~T.~Komiske, E.~M.~Metodiev and M.~D.~Schwartz,
  ``Deep learning in color: towards automated quark/gluon jet discrimination,''
  JHEP {\bf 1701}, 110 (2017)
  [arXiv:1612.01551 [hep-ph]].

\bibitem{Yann:2015nature}
   
   Yann~LeCun, Yoshua~Bengio \& Geoffrey~Hinton,
   ``Deep learning,''
   Nature volume {\bf 521}, 436–444 (2015)


\bibitem{Yoshua:2013ieee}
   
   Yoshua~Bengio, Aaron~Courville, and Pascal~Vincent,
   ``Representation Learning: A Review and New Perspectives,''
   IEEE Transactions on Pattern Analysis and Machine Intelligence {\bf 35}, 8 (2013 )


\bibitem{bjet}
  S.~Chatrchyan {\it et al.} [CMS Collaboration],
  ``Identification of b-quark jets with the CMS experiment,''
  JINST {\bf 8}, P04013 (2013)
  [arXiv:1211.4462 [hep-ex]].

   
\bibitem{Pythia}
  T.~Sjostrand, S.~Mrenna and P.~Z.~Skands,
  ``A Brief Introduction to PYTHIA 8.1,''
  Comput.\ Phys.\ Commun.\  {\bf 178}, 852 (2008)
  [arXiv:0710.3820 [hep-ph]].



   
\bibitem{Fastjet}
  M.~Cacciari, G.~P.~Salam and G.~Soyez,
  ``FastJet User Manual,''
  Eur.\ Phys.\ J.\ C {\bf 72}, 1896 (2012)
  [arXiv:1111.6097 [hep-ph]].

\bibitem{Oleari:2010nx} 
  C.~Oleari,
  ``The POWHEG-BOX,''
  Nucl.\ Phys.\ Proc.\ Suppl.\  {\bf 205-206}, 36 (2010)
  [arXiv:1007.3893 [hep-ph]].
   
   
\bibitem{Delphes}
  J.~de Favereau {\it et al.} [DELPHES 3 Collaboration],
  ``DELPHES 3, A modular framework for fast simulation of a generic collider experiment,''
  JHEP {\bf 1402}, 057 (2014)
  [arXiv:1307.6346 [hep-ex]].
  


   
\bibitem{anti-kt}
  M.~Cacciari, G.~P.~Salam and G.~Soyez,
  ``The Anti-k(t) jet clustering algorithm,''
  JHEP {\bf 0804}, 063 (2008)
  [arXiv:0802.1189 [hep-ph]].
 
   
   
\bibitem{relu}
  
  X~Glorot, A~Bordes, Y~Bengio,
  ``Deep sparse rectifier neural networks,''
  Proceedings of the 14th International Conference on Artificial Intelligence and Statistics, {\bf 15}, 2011  


\bibitem{theano}
  R.~Al-Rfou {\it et al.} [Theano Development Team],
  ``Theano: A Python framework for fast computation of mathematical expressions,''
  arXiv:1605.02688 [cs.SC].

  
  
  

\bibitem{xavier_initialization}
   
   Xavier~Glorot and Yoshua~Bengio,
   ``Understanding the difficulty of training deep feedforward neural networks''
   Proceedings of the 13th International Conference on Artificial Intelligence and Statistics {\bf 9} 2010

\bibitem{adam}
   
   Diederik~P.~Kingma, Jimmy~Ba,
   ``Adam: A Method for Stochastic Optimization,''
   arXiv:1412.6980 [cs.LG]

\bibitem{ensemble_voting}
   
   David~Opitz and Richard~Maclin,
   ``Popular Ensemble Methods: An Empirical Study,'' 
   Journal of Artificial Intelligence Research 11 (1999) 169-198

\bibitem{regularization}
   
   I~Goodfellow, Y~Bengio, and A~Courville,
   ``Deep learning''
   The MIT Press 2016
   
\bibitem{dropout}
  
  Nitish~Srivastava, Geoffrey~Hinton, Alex~Krizhevsky, Ilya~Sutskever, and Ruslan~Salakhutdinov,
  ``Dropout: A Simple Way to Prevent Neural Networks from Overfitting,''
  Journal of Machine Learning Research {\bf 15} 1929-1958 (2014) 
\end{thebibliography}
\end{document}